\documentclass[aps,prl,twocolumn,amsmath,amssymb,nofootinbib,preprintnumbers]
{revtex4-1}

\usepackage{graphicx}
\usepackage{dcolumn}
\usepackage{bm}
\usepackage{amsmath}
\usepackage{amsthm}
\usepackage{color}

\usepackage{amsfonts}
\usepackage{euscript,bbm}
\usepackage{ifthen}
\usepackage{psfrag}
\usepackage{slashed}

\def\ls{\mathrel{\lower4pt\vbox{\lineskip=0pt\baselineskip=0pt
           \hbox{$<$}\hbox{$\sim$}}}}
\def\gs{\mathrel{\lower4pt\vbox{\lineskip=0pt\baselineskip=0pt
           \hbox{$>$}\hbox{$\sim$}}}}
\def\drawbox#1#2{\hrule height#2pt

\hbox{\vrule width#2pt height#1pt \kern#1pt
              \vrule width#2pt}
              \hrule height#2pt}

\def\Asym#1#2{\vcenter{\vbox{\drawbox{#1}{#2}
              \kern-#2pt       
              \drawbox{#1}{#2}}}}

\newcommand{\be}{\begin{equation}}
\newcommand{\ee}{\end{equation}}
\newcommand{\bea}{\begin{eqnarray}}
\newcommand{\eea}{\end{eqnarray}}

\newcommand{\gsim}{\lower.7ex\hbox{$\;\stackrel{\textstyle>}{\sim}\;$}}
\newcommand{\lsim}{\lower.7ex\hbox{$\;\stackrel{\textstyle<}{\sim}\;$}}

\begin{document}

\title{A Rich Tapestry: Supersymmetric Axions, Dark Radiation, and Inflationary Reheating}

\author{Farinaldo S. Queiroz$^{1}$}
\author{Kuver Sinha$^{2}$}
\author{William Wester$^{3}$}

\affiliation{$^{1}$~Department of Physics and Santa Cruz Institute for Particle Physics,\\
University of California, Santa Cruz, CA 95064, USA \\
$^{2}$~Department of Physics, Syracuse University, Syracuse, NY 13244, USA \\
$^{3}$~Fermi National Accelerator Laboratory, Batavia, IL 60510, USA
}

\begin{abstract}
We exploit the complementarity among supersymmetry, inflation, axions, Big Bang Nucleosynthesis (BBN) and Cosmic Microwave Background Radiation (CMB) to constrain supersymmetric axion models in the light of the recent Planck and BICEP results. In particular, we derive BBN bounds coming from altering the light element abundances by taking into account hadronic and electromagnetic energy injection, and CMB constraints from black-body spectrum distortion. Lastly, we outline the viable versus excluded region of these supersymetric models that might account for the mild dark radiation observed.

\end{abstract}
\maketitle


\section{Introduction} 

The Standard Model (SM) is plagued by two unnatural numbers that are yet to be conclusively understood: the strong CP problem and the gauge hierarchy. The former is usually addressed by introducing a spontaneously broken global symmetry (the Peccei-Quinn symmetry) \cite{axionCP,axionCP2}, while the latter is addressed most popularly within the framework of supersymmetry. Presumably, then, the supersymmetric axion finds its place in the framework of particle physics, and the question becomes: what are its consequences? 

The purpose of this work is to highlight the fact that the supersymmetric axion models address the question of dark radiation (interpreted as new degrees of freedom that contribute to the radiation density at the matter-radiation equality) while being $(a)$ tightly constrained by experimental evidence from Big Bang Nucleosynthesis (BBN) and Cosmic Microwave Background (CMB) and $(b)$ constrained, under broad assumptions on inflationary reheating, by the recent observation (to be confirmed) of primordial gravitational waves in the CMB \cite{Ade:2014xna}.
The fact that the framework can address the question of dark radiation is a strong argument for supersymmetric axions.

Eagerly awaited precise measurements of the CMB and Baryonic Accustic Oscilations (BAO) have been reported recently and several combined data analyses have been performed. In particular, the WMAP Collaboration has presented their 9-year data and concluded that, after combining with data from BAO, from the Atacama Cosmology Telescope (ACT), and from the Hubble Space Telescope (HST) ($ H_0 = 73.8\ \pm  2.4\ {\rm km\ s}^{-1}\ {\rm Mpc}^{-1}  $, \cite{HST}), the number of effective massless neutrinos is $N_{eff} = 3.84 \pm 0.4$ \cite{WMAP9R}. More recently, the Planck Collaboration has reported their results, which at face value, does not bias an extra radiation component with $N_{eff} = 3.36 \pm 0.34$ \cite{PLANCKR}. It is important to notice, however, that the Planck collaboration adopted fairly low values (at the $2.2\sigma$ level) for $H_0$ compared to previous studies \cite{HST}. Since $N_{eff}$ and $H_0$ are positively correlated, increasing $H_0$ would naturally yield higher values for $N_{eff}$ \cite{PLANCKR}. Interestingly, adding the $H_0$ measurement and BAO data, the Planck Collaboration  finds, in fact, $N_{eff} = 3.52 ^{+0.48}_{-0.45}$ \cite{PLANCKR}. Lately, the surprising BICEP2 results favor the presence of dark radiation at $3.6\sigma$ when combined studies are performed \cite{bicepscombined}.  

The situation is, however, quite complex and subtle when the combined results of other experiments are considered, even though the recent results clearly favor the presence of dark radiation. In light of the current status regarding dark radiation, we will investigate the conditions under which the supersymmetric axion can accommodate dark radiation while obeying BBN and CMB bounds. 

Dark radiation in our scenario stems from the production of axions via saxion decay. The amount of dark radiation predicted is thus tightly related to the saxion production mechanism, which depends crucially on the inflationary reheating temperature and the saxion mass. Observable tensor modes fix the scale of inflation near the GUT scale, and, under broad assumptions, prefer higher reheat temperatures. Therefore, current results from BICEP2 play an important role in the overall dark radiation-supersymmetric axions setting.

For the benefit of the reader, we summarize our main results here. We find that saxion decay can give the observed value of dark radiation, and the associated physics depends non-trivially on the inflationary reheat temperature. For low reheat $T_r \sim 10^6$ GeV, light saxions with mass below the pion production threshold $\sim 200$ MeV and axion decay constant $f_a \sim 10^{10}$ GeV can give the correct range of $\Delta N_{eff}$ while evading all bounds. For higher reheating temperature $T_r \sim 10^9$ GeV as preferred by observable tensor modes from BICEP2, heavier saxions with mass $\sim \mathcal{O}(10)$ GeV and $f_a \sim 10^{11} - 10^{12}$ GeV that decay before BBN can give the required $\Delta N_{eff}$.

The paper is structured as follows. In Section \ref{sec1} we introduce the supersymmetric axion and our set-up for producing dark radiation. In Section \ref{sec2}, we make the connection to inflation, while in Section \ref{sec3} we give the BBN and CMB bounds. We end with our conclusions.

\section{Supersymmetric Axions and Dark Radiation} \label{sec1}

The supersymmetric hadronic (KSVZ) axion model \cite{KSVXaxion} has been studied extensively; we refer to \cite{Choi:2011yf} for some recent work in other contexts, and extract only the features required for studying dark radiation. The SM fermions are neutral under the PQ symmetry, hence the couplings of SM fermions to the saxion and axion will be ignored. The axion mass is assumed to be negligible (typically $ m_a \sim \mathcal{O}(\mu {\rm eV}) \sim 6 \times 10^{-6}$ eV $(10^{12}$ GeV $/ f_a)$). 

A chiral supermultiplet $S$ containing the axion is introduced. The multiplet may be written as follows
\be
S=  \left(f_a + \frac{s}{\sqrt{2}}\right)\exp\left(\frac{i a}{\sqrt{2}f_a}\right) 
+ \sqrt{2}\theta \psi_a + \theta^2 F^S \,\,, \nonumber
\ee
%
where, as usual, $\theta$ denote the fermionic coordinates.
The scalar component is necessarily complex, consisting of the radial component $s$ (the saxion) and the axionic phase $a$. The superpartner is the axino $\psi_a$.

The interactions of the saxion, axion, and axino with the strong force can be obtained by integrating out the heavy colored fields charged under the PQ symmetry that are introduced in the KSVZ model. In the limit of unbroken supersymmetry, the effective Lagrangian can be obtained from a supersymmetric version of the usual axion interaction term
\be
L \, \supset \, - \frac{\alpha_s}{2\sqrt{2}\pi f_a} \int d^2 \theta S Tr[W^{\alpha} W_{\alpha}] \,\, + \,\, h.c. \,\,\,\,, \nonumber
\ee
where $W_\alpha$ is the vector multiplet containing the gluon field. This sets the Lagrangian for the thermal production of the members of the supermultiplet from the plasma.

On the other hand, the non-thermal production of axions as dark radiation comes from saxion decays, which requires interactions among the members of the supermultiplet. These interactions are obtained from the kinetic energy term of the heavy fields charged under the PQ symmetry. In particular, a field $\Phi_i$ with PQ charge $q_i$ and vev $v_i$ can be expanded as $\Phi_i = v_i exp [q_i (\sigma + i a)/(\sqrt{2}f_a)]$, with a corresponding kinetic term given by
\bea
\label{Axion_interaction}
L^{kin}_{PQ} &=& \sum_{i=1}^N \partial^\mu \phi_i \partial_\mu \phi^{*}_i \, \sim \,  \\
&\sim & \left( 1 + \frac{s}{\sqrt{2}f_a} \right) \left[ \lambda_1(\partial_\mu a)^2 +\frac{\lambda_2}{2} i \overline{\psi_a} \gamma_\mu \partial^{\mu} \psi_a  + \ldots \right]. \nonumber
\eea
Here, $\lambda_1$ and $\lambda_2$ are model-dependent parameters. For the simplest KSVZ model with a single heavy PQ charged field, one has $v = f_a$ and hence $\lambda_1 = \lambda_2 \sim \mathcal{O}(1)$.

%
%
%
%
%



We will take $m_s$ and $m_\psi$ denote the saxion and axino masses, respectively. In gravity-mediated scenarios, the saxion and axino mass are expected to be $\mathcal{O}(m_{3/2})$ due to the presence of supersymmetry breaking effects, in the absence of sequestering or fine-tuning. However, a whole range of axino masses from the eV (produced from a sub-GeV out-of-equilibrium photino) and keV range, to the TeV range, has been considered in the literature. The axino mass $m_{\psi}$ is thus model-dependent; most cosmological studies of the supersymmetric axion assume $m_{\psi}$ to be a free parameter. While we will remain agnostic about the axino mass, we note that its magnitude may have profound implications for the nature of dark matter, about which we will comment later on.

We will similarly assume that the saxion mass is a free parameter, taking values in the $\mathcal{O}($MeV $-$ TeV$)$ range, and that it decays primarily to axions. The decay to axinos may be kinematically blocked, for example, if $m_s \sim m_{\psi}$, or rendered irrelevant, in the limit $m_s \gg m_{\psi}$. In general, the decay widths to axions and gluons are given by \cite{Bae:2013qr},
\bea
\Gamma(s \rightarrow 2a) \, \sim \, \frac{\lambda^2_1}{64 \pi}\frac{m^3_s}{f^2_a} \nonumber \\
\Gamma(s \rightarrow 2g) \, \sim \, \frac{\alpha^2_s}{64 \pi^3}\frac{m^3_s}{f^2_a}
\eea
For $\lambda_1 \sim 1$, $ \mbox{BR}(a) \sim 1$, where $ \mbox{BR}(a)$ is the branching to axions and the branching to gluons satisfies $ \mbox{BR}(g) \, \simeq 10^{-3}  \mbox{BR}(a)$. For $\lambda_1 \sim 10^{-3}$, one has $ \mbox{BR}(a) \sim 10^{-3}$ and $ \mbox{BR}(g) \simeq 1$. The lifetime of the saxion is found to be,
\be \label{DR0}
\tau (s\rightarrow aa) \simeq 1.3\times 10^5  \lambda^{-2}_1 \left( \frac{m_s}{100 \,\, {\rm MeV}} \right) ^{-3} \left( \frac{f_a}{10^{12}\,\, {\rm GeV}} \right)^2 \,\, {\rm sec}.
\ee   
If such decay happens at late times, it surprisingly gives rise to a non-negligible amount of dark radiation in the early universe while circumventing the BBN and CMB constraints on late decays as we shall see further \cite{DRpapers1,DRpapers2}. The idea of dark radiation coming from late decaying particles can be explained simply from Eq.(\ref{rhorad}) below,
\begin{equation}
\rho_{rad} =\left[ 1+ N_{eff} \frac{7}{8} \left(\frac{T_{\nu}}{T_{\gamma}}\right)^4\right] \rho_{\gamma},
\label{rhorad}
\end{equation}
where $T_{\gamma} (T_{\nu})$ is the photon (neutrino) temperature, $N_{eff}$ is the effective number of neutrinos species, and $\rho_{\gamma} = g \pi^2/30 T_{\gamma}^4$ is the photon energy density. 

Notice that after the $e^{\pm}$ annihilation ($T < m_{e}$) the only remaining SM relativistic particles are the CMB photons and the three SM neutrinos. Hence for precisely that reason only the neutrino and photon energy densities have been included in Eq.(\ref{rhorad}). More generally, in addition to the SM neutrinos, there might be extra species that are rather light and weakly coupled contributing to the radiation density or species produced by the decays of heavy particles \cite{DRpapers1,DRpapers2}. The new contribution to the radiation density is often parametrized in terms of the number of neutrinos species $N_{eff}$ as follows \cite{Choi:2012zna},
\be \label{DR2}
 \Delta N_{eff} (t) \,= \, N_{eff,SM} \frac{\rho_{\rm DR}(t)}{\rho_\nu(t)}=  \left(\frac{8}{7}\right) \left( \frac{11}{4}\right)^{4/3}  \frac{\rho_{\rm DR}(t)}{\rho_\gamma(t)},
\ee
where $\rho_\nu=\frac {7}{8} N^{SM}_{eff} T^4_\nu$,$\rho_{\rm DR}$ is the extra relativistic energy density and $\rho_{DR}=\rho_{s}(\tau)=s(\tau)m_s Y_s$, where $s=2\pi^2/45 g_{\star s}T^3$ is the entropy density and $Y_s$ is the saxion abundance. It is worth pointing out that the radiation density is directly related to the saxion abundance, and if all saxions decay into dark radiation then the relation above is preserved. Otherwise one has to introduce the branching ratio to account for the fact that not all saxion decays produce dark radiation, as we will do below.

Notice that the energy densities are written in terms of the temperatures, but there is a time dependence in Eq.(\ref{DR2}). Therefore we need to solve the Friedmann equation for the radiation dominated universe namely,
 $(\dot{T}/T)^2=8 \pi G_N/3 \rho$, where $\rho= \pi^2/30 g_{\star} T^4 $ , in order to find the time-temperature relation and rewrite $\Delta N_{eff}$ in a general form as follows,
 
 \begin{equation}
\Delta N_{eff} (\tau_s) \, = \, 7.1 \left(\frac{\tau_s}{10^6 \, {\rm sec}}\right)^{1/2} \left(\frac{(m_s Y_s)_{tot}}{1 \, {\rm keV}}\right) \mbox{BR}(a) \,\,,
\label{Neff}
\end{equation}
where $Y_s=n_s/s$, with $(m_s Y_s)_{tot}$ accounting for the total thermal and non-thermal saxion production that we will describe further, and $\mbox{BR}(a)$ is the branching ratio into relativistic particles. In Eq.\ref{Neff}, we used the sudden decay approximation.  In this work, we study the case where $ \mbox{BR}(a) = 1$. While we do not compute the case of $\mbox{BR}(a) = 10^{-3}$ explicitly, the BBN bounds are expected to be much stronger there.

Now that we have derived the dark radiation expression in terms of the saxion lifetime, a few remarks are in order:

$(i)$ The entropy and energy densities have different temperature dependences. This is a result of the intrinsic definition of entropy, which is is a measure of the number of specific ways in which a thermodynamic system may be arranged. Consequently, the values of $g_{\star s}$ and $g_{\star}$ which appear in the entropy and energy density are slightly different ($g_{\star s}=3.91$ and $g_{\star}=3.36$), due the difference of the photon and neutrino temperatures caused by the $e^{\pm}$ annihilation. This has been taken into account in the derivation of Eq.(\ref{Neff}). At high temperatures $g_{\star s}$ and $g_{\star}$ are identical though.

$(ii)$ The saxion abundance receives contributions from both thermal and non-thermal processes.  The thermal contribution (produced by scatterings of particles in the high-temperature plasma) depends on the relative values of the inflationary reheating temperature $T_R$ and the temperature $T_{decoup} \sim $ at which the saxion thermally decouples. The thermal abundance is  \cite{Graf:2012hb}, \cite{Rajagopal:1990yx},
\bea  \label{Yth}
(m_s Y_s)_{\rm th} & \sim & 1 \times 10^{-3} \,\,{\rm GeV} \left(\frac{m_s}{1 \, {\rm GeV}}\right)\left(\frac{T_R}{T_{decoup}}\right) \nonumber \\
&&\,\,\,\,\,\,\,\,\,\,\,\,\,\,\,\,\, {\rm for} \,\,\,\,T_R \lsim T_{decoup} \\ \nonumber \\
(m_s Y_s)_{\rm th} & \sim & 1 \times 10^{-3} \,\,{\rm GeV} \left(\frac{m_s}{1 \, {\rm GeV}}\right) \nonumber \\
&&\,\,\,\,\,\,\,\,\,\,\,\,\,\,\,\,\, {\rm for} \,\,\,\,T_R \gsim T_{decoup} \,\,,
\eea
where decoupling temperature is given by $T_{decoup} \sim 10^9 \, {\rm GeV} (f_a/10^{11})^2$.

The non-thermal contribution, produced by coherent oscillations of the saxion field, depends on the relative values of the inflaton decay width $\Gamma_\phi$ and the scale of the onset of the oscillations $H \sim m_s$. Assuming the amplitude of the oscillations is $\sim f_a$, one obtains \cite{Kawasaki:2007mk},
\bea 
(m_s Y_s)_{non-th} & &\simeq 2.2\,\, {\rm GeV} \,\, \left(\frac{T_R}{10^{14} \,\, {\rm GeV}} \right) \left(\frac{f_a}{10^{12} \,\, {\rm Gev}}\right)^2 \nonumber \\ 
& & {\rm for}\,\,\, T_R \lsim \sqrt{m_sM_p}
\label{Ynonth1}
\eea 

\bea 
(m_s Y_s)_{non-th} &\simeq & 0.15 \,\,{\rm GeV} \left(\frac{m_s}{10^{4} \,\, {\rm GeV}} \right)^{1/2} \left(\frac{f_a}{10^{13} \,\, {\rm Gev}}\right)^2 \nonumber \\ 
&&\,\,\,\,\,\,\,\,\,\,\,\,\,\,\,\,\, {\rm for} \,\,\,\,T_R \gsim \sqrt{m_sM_p},
\label{Ynonth}
\eea 
where $M_p$ is the reduced Planck mass. Note that for $m_s \sim 10$ MeV - $1$ GeV, $\sqrt{m_sM_p} \sim 10^9$ GeV.

\begin{figure}[!h]
\centering
\includegraphics[width=\columnwidth]{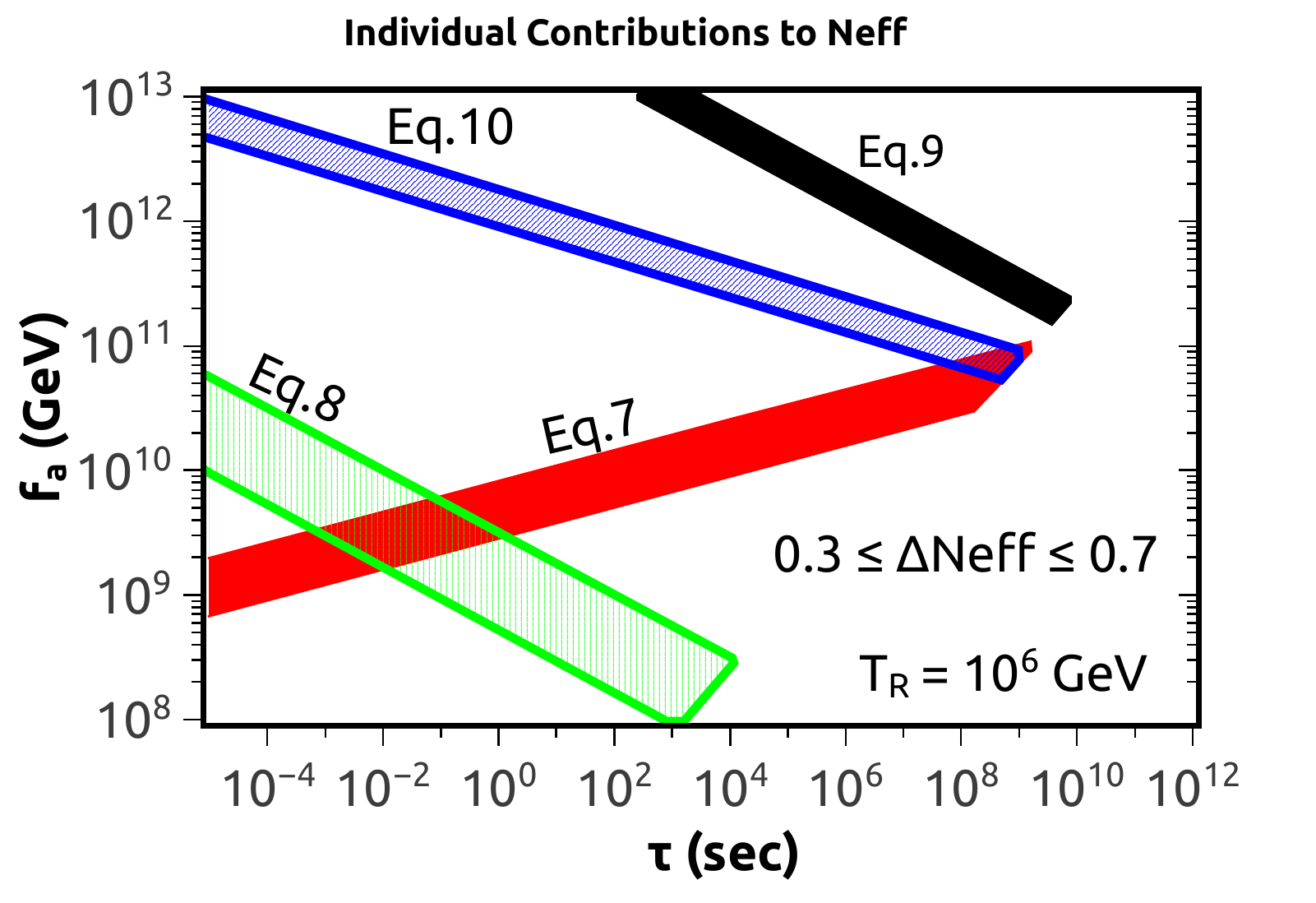}
\caption{Axion decay constant $\times$ lifetime plane that yields $0.3 \leq \Delta N_{eff} \leq 0.7$ for each saxion production setup described in Eqs.(\ref{Yth})-(\ref{Ynonth}) with the saxion mass bounded to be in the MeV-TeV range for $T_R = 10^6$~GeV. We emphasize that we have computed $\Delta N_{eff}$ without forcing the conditions described in Eqs.(\ref{Yth})-(\ref{Ynonth}). The purpose of this plot is to show the how the saxion production mechanisms favor different regions of the parameter space. With this result in mind the reader will be able to easily understand the outcome in Fig.\ref{Graph1} where we accounted for the conditions aforementioned accordingly and exhibited the BBN and CMB bounds.}
\label{Graph0}
\end{figure}

\begin{figure*}[!t]
\centering
\mbox{\hspace*{-1cm}\includegraphics[width=\columnwidth]{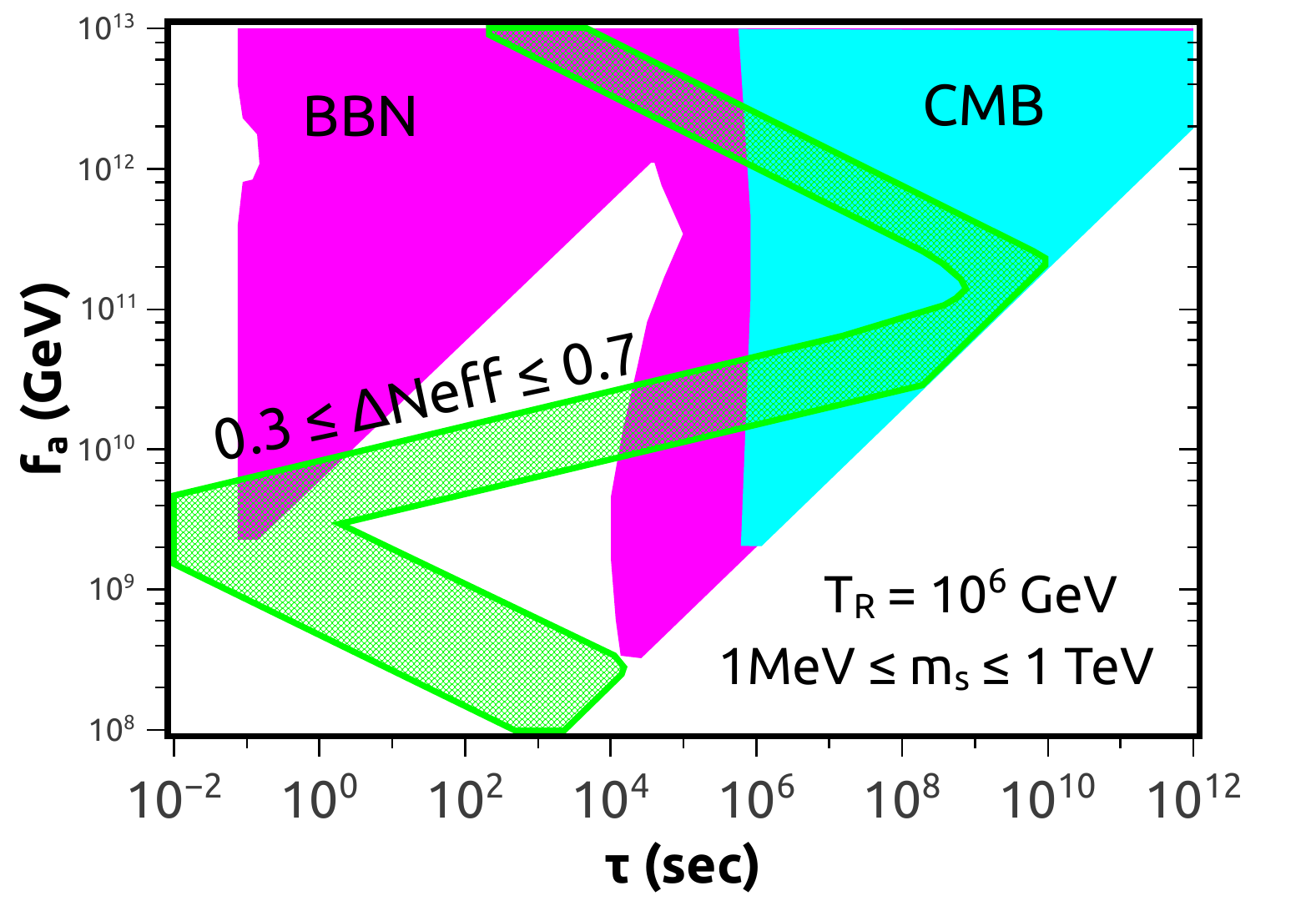}\quad\includegraphics[width=\columnwidth]{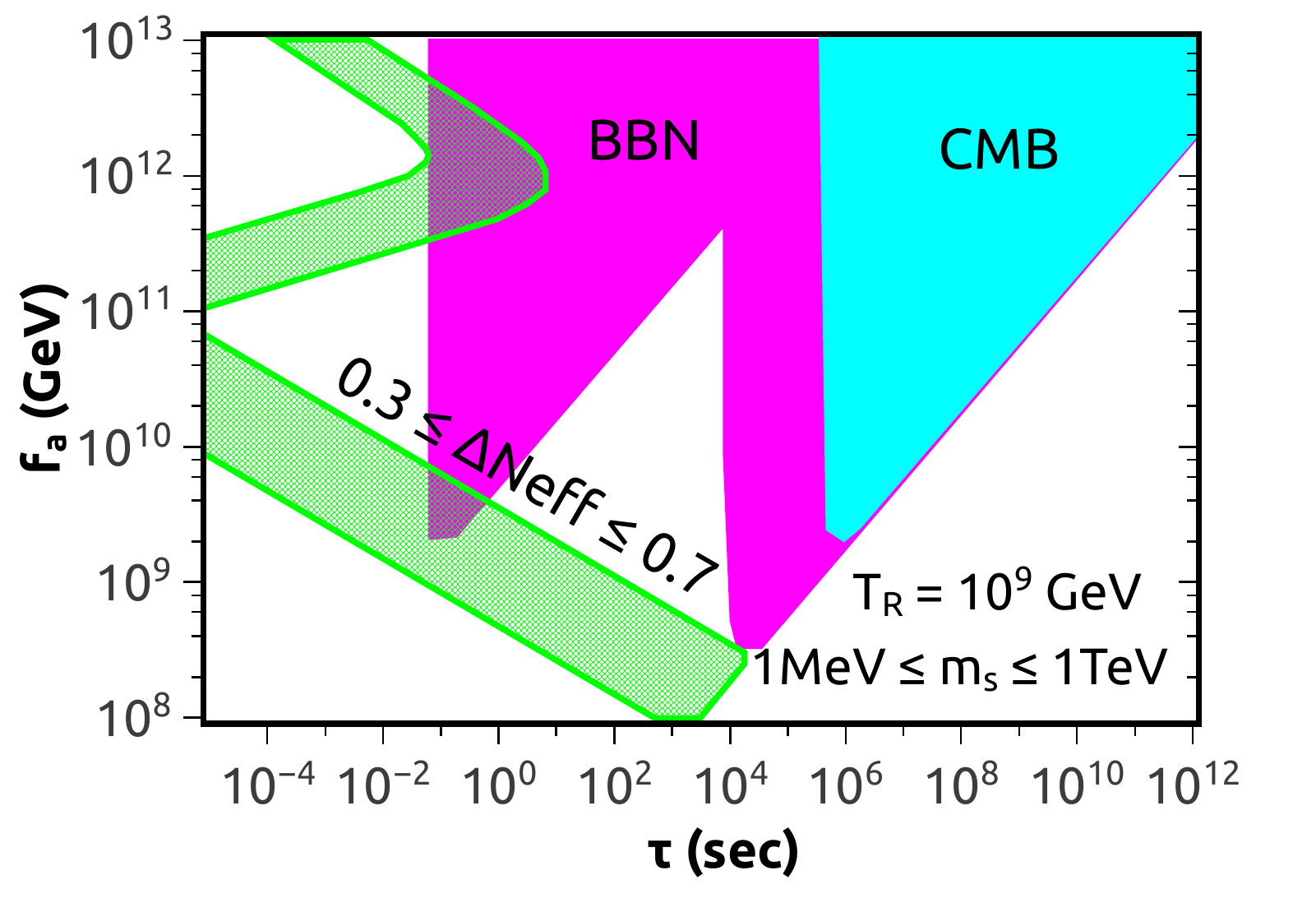}}
\caption{BBN and CMB bounds on the axion decaying constant and lifetime of the saxion for different reheating temperatures. For lifetimes smaller than $10^4$~sec, a simple supersymmetric axion model can accommodate dark radiation and evade the bounds even for high reheating temperatures currently favored by BICEP2 results in $m^2\phi^2$ inflation models.}
\label{Graph1}
\end{figure*}

In Eqs.(\ref{Neff})-(\ref{Ynonth}) we have gathered all the ingredients needed in this dark radiation discussion with $(m_sY_s)_{tot}$ in Eq.(\ref{Neff}) being given by the combination of Eqs.(\ref{Yth})-(\ref{Ynonth}) properly accounting for the reheating temperature conditions as shown above. We note that our expressions agree with the results in the literature; in particular, the expressions for the thermal production agree with the results of \cite{Graf:2012hb} after putting in the relevant value of the strong coupling constant. The non-thermal production depends on the initial amplitude of the saxion field, as we have pointed out; we refer to \cite{Kawasaki:2007mk} for a careful discussion of the production mechanism. Notice that our results depend on three free parameters namely, saxion mass, reheating temperature and axion decay costant. These parameters have a close relation to observables. Hence, we will exploit complementary information coming from different stages of the evolution of the Universe, such as inflation, BBN and CMB to constrain these parameters and draw our conclusions in the next section. For now we will extend our discussions regarding the important physics aspects of this dark radiation and supersymmetric axion setting. We show in Fig.\ref{Graph0}
for $T_R=10^6$~GeV, the axion decay constant $\times$ lifetime plane that yields $0.3 \leq \Delta N_{eff} \leq 0.7$ for each saxion production setup described in Eqs.(\ref{Yth})-(\ref{Ynonth}) with the saxion mass bounded to be in the MeV-TeV range, for saxion-axion branching ratio ($ \mbox{BR}(a)=1$.
In Fig.\ref{Graph0}
we stress that we have computed $\Delta N_{eff}$ without forcing the conditions described in Eqs.(\ref{Yth})-(\ref{Ynonth}). The purpose of this plot is to show the how the saxion production mechanisms induce different regions of the parameter space and that for reasonable  values of lifetime, reheating temperature and axion decaying constant the excess of radiation favored in the current data might be addressed. With the result shown in Fig.\ref{Graph0}
the reader will be able to easily understand the outcome in Fig.\ref{Graph1} where we properly accounted for the conditions aforementioned and presented the BBN and CMB bounds. 

To obtain Fig.1 we took ($m_s,f_a$) pairs and computed the lifetime using Eq.(3). After comparing the decoupling temperature $T_{decoup}$ (given below Eq.(8)) with the reheating temperature, we use either Eq.(7) or Eq.(8) for the saxion abundance. That being determined, we compute $\Delta N_{eff}$ using Eq.(6), and plot the region of the parameter space in the ($f_a,\tau$) plane that yields $0.3 <\Delta N_{eff} < 0.7$ in Fig.\ref{Graph0}. Similarly, for the non-thermal production, given the ($m_s,f_a$) pair, we calculate the lifetime using Eq.(3) and check whether or not $T_R < \sqrt{m_s M_p}$ to select either Eq.(9) or Eq.(10) for the abundance. We then again use Eq.(6) to calculate $\Delta N_{eff}$ and draw the curves.

Note that the contribution to dark radiation from thermal axion production is subdominant to production from saxion decay \cite{Salvio:2013iaa}. Thermally produced axions go out of equilibrium at $T \, > \, m_Z$, where $m_Z$ is the SM $Z$ boson mass, when the number of relativistic degrees of freedom includes all SM particles. Thus, they have a present day temperature of $T \sim T_{\gamma}(g_{*s}/g_{SM})^{1/3}$, which means that they contribute at most $\Delta N^{th}_{eff} \, \sim \, 0.03$ to the number of effective neutrino species.

Moreover, one should in general be careful about entropy diluton from saxion decay \cite{Bae:2013qr}. The dilution factor is given by $max(1,  \mbox{BR}(g)m_s Y_s/T_{decay})$, where $\mbox{BR}(g)$ is the branching to gluons, and $T_{decay}$ is the temperature at which the saxion decays. The values of $T_{decay}$ allowed by BBN and CMB constraints will be evaluated later to be in the keV - MeV range. Moreover, the allowed range of the parameters will typically give $m_s Y_s \lsim 10^{-3}$ GeV. Using $\mbox{BR}(g) \sim 10^{-3}$, we can conclude that the dilution factor from saxion decay is order unity and can be safely ignored in our scenario.

We pause for some comments on dark matter, both thermal and non-thermal, in these models. If a conventional neutralino is the lightest supersymmetric particle (LSP), then dark matter considerations proceed along usual lines\footnote{See Ref.\cite{neutralinobounds} for current indirect detection constraints.}. If the axino, on the other hand, is the LSP, in general there is some model-dependence in the viability of thermal axino dark matter, as well as some dependence on the relative values of the decoupling temperature and the inflationary reheating temperature. In the KSVZ model, for example, axinos are produced thermally in the early Universe through reactions like $q \overline{q} \, \leftrightarrow \, \psi_a \tilde{g}$ with a decoupling temperature of $T_{dec} \sim 10^9 GeV (f_a/10^{11} \mbox{GeV})^2$ and the abundance $\Omega_{\psi_a} h^2 \, \sim \, 0.6 \left(\frac{m_{\psi_a}}{1 \, {\rm keV}}\right)$, in the limit where the inflationary reheating temperature is higher than the decoupling temperature \cite{Choi:2013lwa}. Notice, $\mathcal{O}(keV)$ thermal axinos overclose the Universe in this case and some sort of entropy dilation mechanism is needed to remedy that. The DFSZ axino with mass $\mathcal{O}(keV)$, on the other hand, is a viable warm dark matter candidate for $f_a \sim 10^{10}$ GeV and could be produced by freeze-in. For the scenario investigated here, axinos may be produced non-thermally from saxion decay (if kinematically allowed), with abundance given by $\Omega_{\psi_a} = 2 \mbox{BR}(\psi) (m_{\psi}/m_s) \Omega_{s} \sim 5.4 \times 10^{10}  \mbox{BR}(\psi) (m_{\psi}/100\mbox{GeV}) Y_s$. Depending on the branching $\mbox{BR}(\psi)$ to axinos (which we consider in any case to be $\ll 1$), viable cold or warm axino dark matter may be obtained in these models \cite{Choi:2012zna}.

One should also be careful about the decay of other supersymmetric particles like the gravitino and the next-to-lightest supersymmetric particle (NLSP). The gravitino can decay to the axino and an axion, if kinematically allowed. The amount of dark radiation produced is given by $\Delta N_{eff} \, \sim \, 0.6 (100 \, {\rm GeV}/m_{3/2})^{5/2} (m_{\tilde{g}}/1 \, {\rm TeV})^2 (T_R/10^{10} \, {\rm GeV})$ \cite{Hasenkamp:2011em}. This is negligible for a heavy gravitino, as we assume throughout. Similarly, the NLSP may be a neutralino, and it will decay to the axino LSP, which should happen before BBN. Since the decay is suppressed by the axion decay constant, $f_a$ cannot be too large. The strongest constraint comes in the case of a Bino, which decays to an axino and a photon; the axion decay constant for an $\mathcal{O}(500)$ GeV Bino is constrained to be $\lesssim \, 10^{12}$ GeV \cite{Covi:2001nw}. For Higgsino or Wino NLSP, the bounds are looser \cite{Covi:2009bk}.

We now move on to our main points of interest: inflation, BBN and CMB bounds on dark radiation production.

\section{Connection to Inflation}  \label{sec2}

The connection to inflation comes from the dependence on the reheat temperature in Eq.(\ref{Neff}), which comes through the possible expressions for $(m_s Y_s)_{tot}$ given before. The inflationary reheat temperature will depend on the decay width of the inflaton $T_R = \left(10/g_*(T_R) \pi^2\right)^{1/4}\sqrt{\Gamma_\phi M_p} \,\,.$ The decay width depends on the mass of the inflaton and its decay modes. While the decay modes depend on the specific ways in which the inflaton couples to other fields, the mass of the inflaton can be obtained in a rough estimate from the tensor-to-scalar ratio.

We first turn to the question of decay modes. In many models, the inflaton is a field with gravitational couplings to the particles in the visible sector. For a gravitationally coupled inflaton, the decay width goes as
\be
\Gamma_\phi \, \sim \, \frac{c}{2\pi} ~ \frac{m^3_{\phi}}{M^2_{\rm P}} \,\,.
\label{inflatondecay}
\ee
The precise value of the decay constant $c$ is determined within a UV theory of inflaton decay. While $c$ may take a variety of values, we consider $c \, \sim \, \mathcal{O}(1)$. This is a plausible value; within supergravity, we have explicitly computed the inflaton decay into gauge bosons and gauginos (through a tree-level coupling through the gauge kinetic function), supersymmetric scalars, fermions, and the gravitino (for simple Kahler potentials and superpotentials) summed them all and found Eq.(\ref{inflatondecay}) with $c\sim 1$. Substituting Eq.(\ref{inflatondecay}) into the expression for the reheat temperature we find,
\be \label{reheat2}
T_R \approx 5\times 10^9 \, {\rm GeV} \sqrt{c} \left(\frac{m_\phi}{10^{13}\,{\rm GeV}}\right)^{3/2}
\ee
Now we will use the recent measurements on the tensor-to-scalar ratio to determine the mass of the inflaton. The tensor-to-scalar ratio is defined as $r  \equiv  \Delta^2_t (k)/\Delta^2_s(k)$ with $\Delta^2_t (k) \sim  (2/3 \pi^2) (V/M^4_p) \Delta^2_s (k)  \sim  (1/24 \pi^2) (V/M^4_p) (1/\epsilon)$, where $\epsilon$ is the usual slow-roll paramater, $\Delta^2_t (k)$ and $\Delta^2_s (k)$ are the fluctuation amplitude squared of the tensor and scalar modes, and $k=aH$ is the scale of horizon crossing. Notice that the tensor and scalar perturbations can be expressed in terms of the inflationary potential $V$ and its derivatives in the context of slow-roll inflation. 

The tensor-to-scalar ratio $r$ is related to the scale of inflation, the slow-roll parameter $\epsilon$, and the displacement $\Delta \phi$ of the inflaton $\phi$: $r = 16 \epsilon$; $V^{1/4}  \sim  \left(r/0.01\right)^{1/4} 10^{16} \,\, {\rm GeV};$ $(\Delta \phi/M_p) = \mathcal{O}(1) \times \left(r/0.01\right)^{1/2} \,\,.$

Recent measurements of the WMAP9 satellite point to a scalar spectrum $\Delta^2_s(k) \sim 2.2 \times 10^{-9}$ \cite{WMAP9R} , whereas BICEP2 collaboration has recently announced the detection of primordial cosmic microwave background B-mode polarization, giving a tensor-to-scalar ratio of $r = 0.2^{+0.07}_{-0.05}$ or $r=0.16^{+0.06}_{-0.05}$ after foreground subtraction, as compared to upper bounds from the large-scale CMB temperature power spectrum: $r < 0.13$ (WMAP) or $r < 0.11$ (Planck) at 95\% CL \cite{WMAP9R,PLANCKR}. 


%
For $r = 0.2$, one obtains $\epsilon \, \sim \, 1.3 \times 10^{-2}$ and $\Delta \phi \, \sim \, \mathcal{O}(M_p)$. For the case of $V = m^2 \phi^2$, this further implies $m_{\phi} \, \sim \, 10^{12} \,\, {\rm GeV} \,\,.$

In fact, large field models with measurable tensor-to-scalar ratio generally prefer the inflaton mass $\sim 10^{12}$ GeV. One thus obtains, combining Eqs.~\ref{reheat2} and the above discussion, that the reheat temperature corresponding to the observed tensor-to-scalar ratio is
\be \label{reheatfinal}
T_R \, \sim \, 10^9 \,\, {\rm GeV} \,\,.
\ee
Notice that we have brought information from cosmological inflation to pin down the favored reheating temperature due to current measurements. This argument relies, of course, on broad assumptions; a full theory of reheating and pre-heating, which would include a precise knowledge of the inflaton Lagrangian, would be needed to make more precise statements.

Given the uncertainties concerning BICEP2 results and the fact that different values of $r$ imply different reheating temperatures, we will draw our conclusions for different reheating temperatures ($T_R=10^6,10^9$~GeV), merely noting that  Eq.(\ref{reheatfinal}) is the value preferred by current observation, under broad assumptions.

Now we will connect our setup of supersymmetric axions as dark radiation with BBN and CMB observables.
 
\section{Connection to BBN and CMB} \label{sec3}

The light elements abundances and the power spectrum of the cosmic microwave brackground radiation are the first places one should look at the for the presence of new light species in the early universe given the current precise measurements on both observables. Since we are advocating the presence of late decays of saxions (mother particles) into axions/gluons (daughter particles), stringent constraints can be placed on the parameters that set the lifetime and energy injection in this setup, because those are the parameters that BBN and CMB are most sensitive to for a given final state decay. That being said, we follow the procedure described in Ref.\cite{Kawasaki:2004qu} which used a modified version of the KAWANO code \cite{kawanocode} to compute the creation and destruction of the light elements such deuterium (D), $^4$He, $^6$He,$^7$Li, due to electromagnetic and hadron showers induced by the energy injection produced in the saxion decays. We have used in our calculations the bounds, $3.31\times 10^{-5} < D/H <4.57 \times 10^{-5}; 0.227 <Y_p (^4He) < 0.249; 1.09 \times 10^{-10} < ^7Li/H < 4.39 \times 10^{-10}$ and $0.18 \times 10^{-11} < ^6Li/H < 6.1 \times 10^{-11}$ \cite{Kawasaki:2004qu}. Because in our scenario the saxion has a sizable branching ratio into gluons ($10^{-3}$) the most stringent constraint tipicaly comes from the
hadrodissociation, which is induced by the emission of energetic neutron and proton.  Although, if the production of the nuclei is kinematically blocked, hadronic constraints
become weak. Albeit, the saxion decays with a large branching ratio  (close to unity) into axions that can be converted into photons, which in turn induce photo-dissociation ( $\gamma + X \rightarrow Y$) capable of still significantly altering the abundances in concern. We have taken these effects into account as well following the procedure described in  Ref.\cite{BBNbounds}. In summary, BBN constrains $(m_sY_s)_{tot}$ for a given lifetime. This bound can be translated into limits in the axion decaying constant  for a given reheating temperature using Eqs.\ref{Yth}-\ref{Ynonth}.

Additionally, we derived CMB bounds related to distortions to the CMB black-body spectrum due to the energy injection. In this case, the photon spectrum relaxes to a Bose-Einstein distribution with a chemical potential different from zero. Limits on chemical potential, namely $\mu < 9 \times 10^{-5}$, can then be used to constrain this additional energy injection as follows \cite{HuR},
\begin{equation}
\mu = 8 \times 10^{-4} \left(\frac{\tau}{10^6 sec}\right) \left( \frac{ (m_s Y_s)_{tot} }{2 10^{-9}}\right) \exp^{ -(t_c/\tau)^{5/4}}
\end{equation}where
\begin{equation}
t_c=6.1 \times 10^6 \mbox{sec} \left(\frac{T_0}{2.725K}\right)^{-12/5}\left(\frac{\Omega_b h^2}{0.022}\right)^{4/5}\left(\frac{1-Y_p/2}{0.8}\right)^{4/5}
\end{equation}

We now turn to a discussion of the results from Fig.~\ref{Graph1}. The pink region is excluded by BBN, whereas blue one reflects the CMB constraints following the recipe above. The green area shows the region of parameter space that reproduces $0.3 \leq \Delta N_{eff} \leq 0.7$. We display our results for two cases: inflationary reheat temperature of $T_R \sim 10^6$ GeV and $T_R \sim 10^9$ GeV. The allowed regions are shown on the plane of saxion lifetime $\tau$ and the axion decay constant $f_a$; thus, every point on the plane corresponds to a different saxion mass determined from Eq.(\ref{DR0}). The saxion mass is varied from 1 MeV to 1 TeV. We conclude that BBN is sensitive to smaller and large lifetimes, whereas CMB is just sensitive to larger lifetimes. The latter has to do with the thermalization processes which are efficient at higher temperatures, leaving no imprint on the CMB.

Although we used exact bounds in our calculations, it is useful to mention order-of-magnitude approximations to better understand the figures. Assuming a branching to hadrons $\sim \mathcal{O}(10^{-3})$, the energy density of the decaying particle is approximately constrained to be $m_s Y_s \lesssim 10^{-11}$ GeV for a lifetime of order of $10^5$ sec. This is smaller than the $(m_s Y_s)_{tot}$ obtained from Eq.(\ref{Yth})-(\ref{Ynonth}) for a large region of the parameter space; hence these regions are excluded. The triangular allowed wedge corresponds to $m_s \, < \, 200$ MeV, for which the hadronic constraints basically vanish, whereas the electromagnetic ones still give rise to some important bounds.

It is clear from Eq.(\ref{Yth})-(\ref{Ynonth}) that smaller the reheat $T_R$, smaller the quantity $(m_s Y_s)_{tot}$. Consequently, from Eq.~\ref{Neff}, to obtain a fixed $\Delta N_{eff}$, smaller $T_R$ requires greater lifetime $\tau$. This is why the green allowed region for $\Delta N_{eff}$ is at higher lifetimes for the case $T_R = 10^{6}$ GeV. 

It is also instructive to understand the shape of the $\Delta N_{eff}$ green regions. For simplicity, we consider the case $T_R = 10^9$ GeV. From Eq.(\ref{Yth})-(\ref{Ynonth}), it is clear that $(i)$ for large $f_a \gsim 10^{12}$ GeV, the thermal contribution is suppressed by the ratio $T_R/T_{decoup}$, and the non-thermal contribution dominates; as $f_a$ becomes smaller in this region, $(m_s Y_s)_{tot}$ decreases and $\tau$ must increase to keep $\Delta N_{eff}$ in the allowed region, $(ii)$ for $f_a \lesssim 10^{11}$ GeV, the non-thermal contribution becomes small, while the thermal contribution becomes maximal, and the allowed region of $\Delta N_{eff}$ shows the same trend as above, and $(iii)$ in the region $10^{11}$ GeV $\lesssim \, f_a \, \lesssim \, 10^{12}$ GeV, the two contributions can become comparable; for $m_s \gsim \mathcal{O}(1)$ GeV, the thermal contribution can dominate and the saxion decays early to produce the same amount of dark radiation. This has been clearly shown in Fig.\ref{Graph0} previously.


We will mainly be interested in the region $10^{10}$ GeV $\leq \, f_a \, \leq \, 10^{12}$ GeV. For the case of $T_R \sim 10^6$ GeV, the green $\Delta N_{eff}$ region is almost entirely ruled out, except for some regions with $f_a \sim 10^{10}$ GeV which evade hadronic and electromagnetic energy injection bounds. 

{\bf Summary of our Results} 

Higher saxion masses are preferred in the higher reheat regime by BBN constraints. As discussed above, the green region corresponding to the allowed range of $\Delta N_{eff}$ moves to the left (shorter saxion lifetime or, equivalently, large saxion mass) as the reheat temperature is increased. In our Figure 2, for $T_r = 10^9$ GeV, we similarly see that the allowed green band  lies mainly in the small saxion lifetime $\tau$ region, corresponding to higher saxion mass, while the large saxion lifetime (small saxion mass) regions start conflicting with BBN bounds. Our results are in agreement with \cite{Graf:2012hb}. 

{\it Result 1}: For low reheat temperatures, $T_R \sim 10^6$ GeV, and saxion-axion branching ratio close to unity, the supersymmetric axion can accommodate the observed dark radiation for light saxionic decay with $m_s \sim 200$ MeV, and $f_a \sim 10^{10}$ GeV as shown in Fig.\ref{Graph1}. 

{\it Result 2}: For high reheat temperatures, $T_R \sim 10^9$ GeV, and saxion-axion branching ratio close to unity, the green curve has shifted to the left, and there is a large area that evades the BBN and CMB constraints entirely. In this region, the saxion is massive and decays at higher temperatures. As a benchmark point, the choices $f_a \sim 10^{11}-10^{12}$ GeV and $m_s \sim 10$ GeV reproduce $\Delta N_{eff} \sim 0.5$. 

Therefore, for high reheat temperatures as preferred by the current observed tensor modes, the supersymmetric axion can accommodate the observed dark radiation.

\section {Conclusions} 

In this work we have exploited the complementarity among axions, inflation, BBN and the CMB radiation in the context of supersymmetry in the wake of the recent results from Planck and BICEP2 regarding dark radiation and the observation of B-mode polarization in the CMB.

First, we derived the expression for dark radiation component resulted from saxion decays and connected it to different saxion production mechanisms which are comprised of thermal and non-thermal production and depend on various input such as the saxion mass, reheating temperature and axion decay constant.

Further, we brought inflation inputs into our analysis, which tells us that a reheating temperature of $10^9$~GeV is preferred in the vanilla inflation model. Later, we derived BBN and CMB bounds coming from the alteration of the light element abundances and deviations from the Black-body spectrum respectively, in the axion decay constant $\times$ saxion lifetime plane, taking into account hadronic and electromagnetic energy injection. We presented those constraints in the regime where branching ratio into electromagnetic energy injection is close to unity. Lastly, we combined all this information to outline the viable versus excluded region of the parameter space that might account for the mild dark radiation observed, while obeying the existing bounds aforementioned. We have found two benchmark models satisfying all bounds,

(i) {\it Low Reheat: } The supersymmetric axion can accommodate the observed dark radiation for saxion mass $m_s \sim 200$ MeV, and axion decay constant $f_a \sim 10^{10}$ GeV.

(ii) {\it High Reheat: } The supersymmetric axion can reproduce the observed dark radiation for $m_s \sim 10$ GeV, and $f_a \sim 10^{11} - 10^{12}$ GeV.

The trend shows that higher saxion masses are preferred in the higher reheat regime by BBN constraints. 

 \section{Acknowledgements}

The authors are indebted to Takeo Moroi and Kazunori Kohri for clarifications regarding BBN constraints. The authors would like to thank Tom Banks, Alex Dias, Michael Dine, Patrick Draper, Jiji Fan, Kazunori Kohri, Takeo Moroi, Ogan \"Ozsoy, Carlos Pires, Paulo Rodrigues, William Shepherd, and Scott Watson for useful discussions. The authors thank the organizers of the Mitchell Workshop in Texas where this project was initiated. FQ is partly supported by US Department of Energy Award SC0010107 and the Brazilian National Counsel for Technological and Scientific Development (CNPq). KS is supported by NASA Astrophysics Theory Grant NNH12ZDA001N. WW is supported by the Fermi Research Alliance, LLC under Contract No DE-AC02-07CH11359 with the United States Department of Energy. 

\appendix

\end{document}